\documentclass[conference]{IEEEtran}
\usepackage[applemac]{inputenc}
\usepackage[english]{babel}

\usepackage{amsmath}
\usepackage{amssymb}
\usepackage{amsthm}
\usepackage{thmtools}

\newtheorem{lemma}{Lemma}
\newtheorem{theorem}{Theorem}
\newtheorem{corollary}{Corollary}

\newtheoremstyle{mydef}
	{3pt}		
	{3pt}		
	{}		
	{}		
	{\itshape}	
	{:}		
	{.5em}	
	{}		

\theoremstyle{mydef}

\usepackage{cite} 

\ifCLASSOPTIONcompsoc
	\usepackage[caption=false,font=normalsize,labelfont=sf,textfont=sf]{subfig}
\else
	\usepackage[caption=false,font=footnotesize]{subfig}
\fi

\usepackage{tikz}
\usetikzlibrary{calc,
shapes,
positioning,
snakes,
patterns}

\makeatletter
\tikzset{
        hatch distance/.store in=\hatchdistance,
        hatch distance=5pt,
        hatch thickness/.store in=\hatchthickness,
        hatch thickness=5pt
        }
\pgfdeclarepatternformonly[\hatchdistance,\hatchthickness]{north east hatch}
    {\pgfqpoint{-1pt}{-1pt}}
    {\pgfqpoint{\hatchdistance}{\hatchdistance}}
    {\pgfpoint{\hatchdistance-1pt}{\hatchdistance-1pt}}%
    {
        \pgfsetcolor{\tikz@pattern@color}
        \pgfsetlinewidth{\hatchthickness}
        \pgfpathmoveto{\pgfqpoint{0pt}{0pt}}
        \pgfpathlineto{\pgfqpoint{\hatchdistance}{\hatchdistance}}
        \pgfusepath{stroke}
    }
\makeatother

\usetikzlibrary{plotmarks}
\usepackage{pgfplots}

\newcommand{\gBS}{\gamma_\text{\textnormal{BS}}}
\newcommand{\gD}{\gamma_\text{\textnormal{D2D}}}

\newcommand{\nrBS}{n_{\text{\textnormal{r}}}^\text{\textnormal{BS}}}
\newcommand{\nrD}{n_{\text{\textnormal{r}}}^\text{\textnormal{D2D}}}

\newcommand{\rhoBS}{\rho_\text{\textnormal{BS}}}
\newcommand{\rhoD}{\rho_\text{\textnormal{D2D}}}

\newcommand{\aMBR}{\alpha_\text{MBR}}
\newcommand{\aMDS}{\alpha_\text{MDS}}
\newcommand{\aMSR}{\alpha_\text{MSR}}
\newcommand{\arep}{\alpha_\text{rep}}
\newcommand{\bMBR}{\beta_\text{MBR}}
\newcommand{\bMDS}{\beta_\text{MDS}}
\newcommand{\bMSR}{\beta_\text{MSR}}
\newcommand{\brep}{\beta_\text{rep}}

\newcommand{\dr}{\Delta}
\newcommand{\drmax}{\Delta_\text{max}}

\newcommand{\kc}{k}
\newcommand{\dacc}{h}
\newcommand{\nc}{n}
\newcommand{\Rc}{R}

\newcommand{\Cd}{C_\text{\textnormal{d}}}
\newcommand{\Cr}{C_\text{\textnormal{r}}}

\newcommand{\EC}{\mathbb{E}(C)}
\newcommand{\ECd}{\mathbb{E}\left(\Cd\right)}
\newcommand{\ECr}{\mathbb{E}\left(\Cr\right)}

\newcommand{\PrBSd}{\Pr\{\text{\textnormal{BS download}}\}}
\newcommand{\PrBSds}{\Pr\{\text{\textnormal{BS down.}}\}}

\newcommand{\PrDd}{\Pr\{\text{\textnormal{D2D download}}\}}
\newcommand{\PrDds}{\Pr\{\text{\textnormal{D2D down.}}\}}

\newcommand{\Ta}{T_\text{a}}
\newcommand{\Tl}{T_\text{l}}
\newcommand{\Tr}{T_\text{r}}

\IEEEoverridecommandlockouts

\title{Repair Scheduling in Wireless Distributed Storage with D2D Communication}

\begin{document}

\author{
\IEEEauthorblockN{Jesper Pedersen$^\dag$, Alexandre Graell i Amat$^\dag$, Iryna Andriyanova$^\ddag$, and Fredrik Br\"annstr\"om$^\dag$}
\IEEEauthorblockA{$\dag$Department of Signals and Systems, Chalmers University of Technology, Gothenburg, Sweden\\
  $\ddag$ETIS Laboratory, ENSEA/University of Cergy-Pontoise/CNRS, Cergy-Pontoise, France}\thanks{This work was partially funded by the Swedish Research Council under grant \#2011-5961.}\vspace{-3ex}}

\maketitle

\begin{abstract}
We consider distributed storage (DS) for a wireless network where mobile devices arrive and depart according to a Poisson random process. Content is stored in a number of mobile devices, using an erasure correcting code. When requesting a piece of content, a user retrieves the content from the mobile devices using device-to-device communication or, if not possible, from the base station (BS), at the expense of a higher communication cost. We consider the repair problem when a device that stores data leaves the network. In particular, we introduce a repair scheduling where repair is performed (from storage devices or the BS) periodically. We derive analytical expressions for the overall communication cost of repair and download as a function of the repair interval. We illustrate the analysis by giving results for maximum distance separable codes and regenerating codes. Our results indicate that DS can reduce the overall communication cost with respect to the case where content is only downloaded from the BS, provided that repairs are performed frequently enough. The required repair frequency depends on the code used for storage and the network parameters. In particular, minimum bandwidth regenerating codes require frequent repairs, while maximum distance separable codes  give better performance if repair is performed less frequently. We also show that instantaneous repair is not always optimal.
\end{abstract}


\section{Introduction}
It is predicted that global mobile data traffic will reach 24.3 exabytes per month by 2019, nearly a tenfold increase compared to the traffic in 2014 \cite{Cisco2011}. 
This dramatic increase in mobile data traffic threatens to completely congest the already burdened wireless networks. One popular approach to reduce peak traffic is to store popular data closer to the end users, a technique also known as \textit{caching}. Recently, a novel architecture was proposed to efficiently handle highly predictable bulky traffic, such as video traffic \cite{Shanmugam2013}. The idea is to deploy a number of access points (called helpers) with large storage capacity, but low-rate wireless backhaul, and store data across them. Users can then download content from the helpers, resulting in a performance gain. 

In \cite{Golrezaei2014} it was suggested to store content directly in the mobile devices, taking advantage of the high storage capacity of modern smart phones and tablets. Hence, no additional infrastructure is required. Traffic to the BS is alleviated by maximizing the number of times a requested file can be retrieved from the mobile devices storing content, using device-to-device (D2D) communication. The problem of repairing the lost data when a device leaves the network was considered in \cite{6825065}, where data is stored in the mobile devices using erasure correcting coding. In particular, the communication cost incurred by data download and repair is analyzed in \cite{6825065}, assuming an infinite storage capacity in the mobile devices and instantaneous repair.



In this paper, we consider distributed storage (DS) in a wireless network scenario similar to the one in \cite{6825065}. We consider a cellular system where mobile devices roam in and out of a cell according to a Poisson random process and request content at random times. The cell is served by a base station (BS), which always has access to the content. Content is also stored across a limited number of mobile devices using an erasure correcting code. When a user requests a piece of content, it attempts to download it from the mobile devices using D2D communication. If not possible, the content is downloaded from the BS, at the expense of a higher communication cost. Our main focus is on the repair problem when a device that stores data leaves the network. In particular, we introduce a repair scheduling where lost content is repaired (from storage devices sojourning in the cell or from the BS) at periodic times. We derive analytical expressions for the total communication cost of repair and download as a function of the repair interval. Furthermore, we analyze several erasure correcting codes, namely maximum distance separable (MDS), and regenerating codes. We show that DS can reduce the overall communication cost as compared to the classical scenario where content is only downloaded from the BS, provided that repairs are performed frequently enough. The required frequency depends on the code family and on the network parameters. Somewhat surprisingly, instantaneous repair is not always the optimal.




\section{System Model}\label{sec:SystemModel}

\begin{figure}[!t]
\centering
\includegraphics{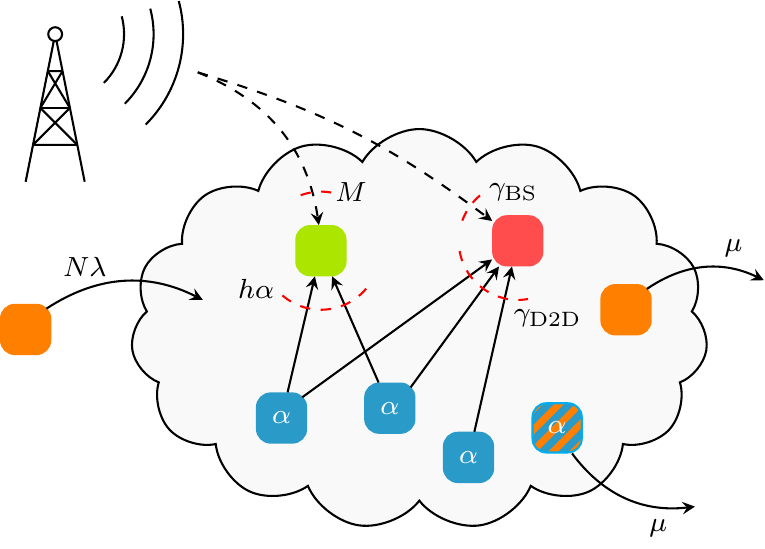}
\vspace{-2ex}
\caption{A wireless network with data storage in the mobile devices (nodes). A new node arrives to the network at rate $N \lambda$. The departure rate per node is $\mu$. Blue nodes store exactly $\alpha$ bits each. The green node requests the file and downloads it from the storage nodes (solid arrows), or from the BS (dashed arrow). The repair of a node (in red) is carried out by transmitting $\gD$ bits from storage nodes (solid arrows) or $\gBS$ bits from the BS (dashed arrow).
	}\label{fig:sys}
	\vspace{-2ex}
\end{figure}

We consider a single cell in a cellular network, served by a BS, where mobile devices (referred to as nodes) arrive and depart according to a Poisson process. The average number of nodes in the network is $N$. Nodes wish to download content from the network. For simplicity, we assume that there is a single object (file), of size $M$ bits, stored at the BS. We further assume that nodes can store data and communicate between them using D2D communication. The considered scenario is depicted in Fig.~\ref{fig:sys}. 

{\it Arrival-departure model.} 
Nodes arrive according to a Poisson process with exponential independent, identically distributed (i.i.d.) random inter-arrival times $\Ta$ with probability density function (pdf)
\begin{equation}\label{eq:Tapdf}
f_{\Ta}(t)=N\lambda e^{-N\lambda t},\quad t\geq0,
\end{equation}
where $N\lambda$ is the expected arrival rate of a node and $t\in\mathbb{R}$ is time, measured in time units (t.u.).

The nodes stay in the cell for an i.i.d. exponential random lifetime $\Tl$ with pdf
\begin{equation}\label{eq:Tlpdf}
f_{\Tl}(t)=\mu e^{-\mu t},\quad t\geq0,
\end{equation}
where $\mu$ is the expected departure rate of a node. The number of nodes in the cell can be described by an $\text{M}/\text{M}/\infty$ queuing model. We assume that $\mu=\lambda$, i.e., the average number of  nodes in the cell stays constant (equal to $N$). 

%
%
%
%
%
%


{\it Data storage.}
The file is partitioned into $\kc$ packets and encoded using an $(\nc,\kc)$ erasure correcting code of rate $\Rc=\kc/\nc$. The encoded data is stored in $n$ nodes, referred to as \emph{storage nodes}. For simplicity, we assume $n\ll N$, hence the probability that the number of nodes in the cell is smaller than $n$ is negligibly small. 
Therefore, the file can always be stored in the network. In particular, each storage node stores exactly $\alpha$ bits, i.e., we consider a symmetric allocation \cite{Leong2012}. Hence,
\begin{align}
\alpha =\frac{M}{\kc}.
\label{eq:alpha}
\end{align}
Like \cite{Leong2012}, we also introduce an overall storage budget constraint of $\Gamma M$ bits, $\Gamma\ge1$, across the nodes in the cell, i.e., $n\alpha\leq \Gamma M$. Note that to satisfy this constraint, $\Rc\ge 1/\Gamma$. 


{\it Data delivery.}
Nodes request the file at random times with i.i.d. random inter-request time $\Tr$ with pdf
\begin{equation}\label{eq:Trpdf}
	f_{\Tr}(t)=\omega e^{-\omega t},\quad t\geq0,
\end{equation}
where $\omega$ is the expected request rate per node. 
Whenever possible, the file is downloaded from the storage nodes using D2D communication, referred to as D2D download. In particular, we assume that data can be downloaded from any subset of $\dacc\in\{k,\ldots,n\}$ storage nodes. In other words, D2D download is possible if $\dacc$ or more storage nodes remain in the cell. In this case, the amount of downloaded data is $\dacc\alpha\ge M$ bits, where the inequality follows because $\dacc\ge k$. The parameter $\dacc$ depends on the properties of the erasure correcting code used for storage, and will be discussed in Section~\ref{sec:codes}. In the case where there are less than $\dacc$ storage nodes in the cell, the file is downloaded from the BS, referred to as BS download. In this case, $M$ bits are downloaded. To simplify the analysis in Section~\ref{sec:analysis}, we assume that the download bandwidth is the same irrespective of whether the request comes from a storage node itself or not. This is a reasonable approximation, since $n\ll N$. 

We assume that transmission from the BS and from a node (in D2D communication) have different costs. We denote by $\rhoBS$ and $\rhoD$ the cost (in cost units (c.u.) per bit, [c.u./bit]) of transmitting one bit from the BS and from a node, respectively, and by $\rho=\rhoBS/\rhoD$ its ratio. We further assume $\rho\ge 1$, hence transmission from the BS is at least as costly as the transmission in D2D communication. 


\subsection {Repair Process}

When a storage node leaves the network, its stored data is lost (see blue node with orange stripes in Fig.~\ref{fig:sys}). Therefore, 
another node needs to be populated with data to maintain the initial state of reliability of the DS network, i.e., $n$ storage nodes. The restore (repair) of the lost data onto another node, chosen uniformly at random from all nodes in the cell that do not store any content, will be referred to as the repair process. In particular, we introduce a scheduled repair scheme where the repair process is launched periodically. We denote the interval between two repairs by $\dr$ (in t.u.), $\dr \ge 0$. Note that $\dr=0$ corresponds to the case of instantaneous repair, considered in \cite{6825065}.

Similarly to the download, repair can be accomplished from the storage nodes (D2D repair) or from the BS (BS repair), with cost per bit $\rhoD$ and $\rhoBS$, respectively. The amount of data (in bits) that needs to be retrieved from the network to repair a single failed node is referred to as the repair bandwidth, $\gamma$. In particular, we assume that D2D repair can be performed from any subset of $r\in\{k,\ldots,n-1\}$ storage nodes by retrieving $\beta\le\alpha$ bits from each node. In other words, D2D repair is possible if there are at least $r$ storage nodes in the cell at the moment of repair. $r$ is usually referred to as the repair access in the literature. In this case $\gD=r\beta$, where the subindex indicates that repair is performed from the storage nodes. If there are less than $r$ storage nodes in the network at the moment of repair, then the repair is carried out by the BS. In this case $\gBS=\alpha$. We assume that repair always succeeds. Furthermore, for both repair and download we assume error-free transmission.






\section{Repair and Delivery Cost}
\label{sec:analysis}


In this section, we derive analytical expressions for the repair cost, $\ECr$, download cost, $\ECd$, and total cost $\EC=\ECr+\ECd$, as a function of the repair interval, $\dr$. The cost is defined in cost units per bit and time unit [c.u./(bit$\times$t.u.)]


\subsection{Average Repair Cost}

Denote by $\nrD$ and $\nrBS$ the average number of nodes repaired from the storage nodes and from the BS, respectively, in one repair interval. Also, let $\{b_i(n,p)\}_{i=0}^n$ be the probability mass function (pmf) of the binomial distribution with parameters $n$ and $p$.
\begin{lemma}
\label{lemma:def-Cr}
\begin{align}
\label{eq:nrd2d}
\nrD   & =  \sum_{i=r}^{n}(n-i) b_i(n, p),\\
\label{eq:nrbs}
\nrBS & = \sum_{i=0}^{r-1}(n-i) b_i(n, p),
\end{align}
where $p= e^{-\mu\dr}$.
\end{lemma}
\begin{IEEEproof}
As the inter-departure times are exponentially distributed, the probability that a storage  node has not left the network during a time $\dr$ and is accessible for repair is $p=e^{-\mu\dr}$.   
Hence, the probability that $i$ storage nodes are accessible is $b_i(n,p)$. If only $i$ storage nodes remain in the network, then $n-i$ repairs need to be performed. D2D repair is performed if $i \ge r$; BS repair is performed otherwise. 
Therefore, (\ref{eq:nrd2d}) and (\ref{eq:nrbs}) hold.
\end{IEEEproof}

The average repair cost, $\ECr$, is given in the following theorem.
\begin{theorem}\label{thm:Cr}
Consider the DS network in Section~\ref{sec:SystemModel} with parameters $M$, $\dr$, $\rhoBS$, $\gBS$, $\rhoD$, $\gD$, $\mu$, $n$ and $r$. The average repair cost is
\begin{align}
\label{eq:Cr}
\ECr&=\frac{1}{M\dr}\left(\rhoBS \gBS \nrBS +\rhoD\gD\nrD\right)\\
&=\frac{1}{M\dr}\left(\rhoBS \gBS \sum_{i=0}^{r-1}(n-i) b_i(n, p)\right.\nonumber\\
\label{eq:Cr2}
&~~~~~~~~~~~~~\left. + \rhoD\gD \sum_{i=r}^{n}(n-i) b_i(n, p)\right),
\end{align}
where $p= e^{-\mu\dr}$.	
\end{theorem}

\begin{IEEEproof}
From the system model, it follows that the cost of repairing a single storage node from the BS is $\rhoBS \gBS$ c.u.
Similarly, the cost of D2D repair of a single node  is $\rhoD\gD$ c.u..
Normalizing by the file size ($M$ bits) and the duration of the repair interval $\dr$, we obtain (\ref{eq:Cr}) in [c.u./bit$\times$t.u.]. Finally, using Lemma~\ref{lemma:def-Cr}, we obtain (\ref{eq:Cr2}).
\end{IEEEproof}

\subsection{Average Download Cost}
\label{sec:delivery-computation}
The average download cost is given in the following theorem.
\begin{theorem}\label{thm:Cd} 
Consider the DS network in Section~\ref{sec:SystemModel} with parameters $N$, $\omega$, $M$, $\rhoBS$, $\rhoD$, $n$, $\dacc$, $\alpha$, $\mu$ and $\dr$. Let $\mu_i=i\mu,$
for $i\in\{\dacc,\ldots,n\}$, and $p_i = e^{-\mu_i\dr}$. 
Then
{\small
\begin{align}
&\ECd=\frac{N\omega}{M} \left(\rhoBS M \PrBSds
		+\rhoD \dacc\alpha\PrDds\right)\nonumber\\
		&=N\omega\Bigg(\rhoBS
			+\left(\rhoD\frac{\dacc\alpha}{M}-\rhoBS\right)\frac{1}{\dr}\sum_{i=\dacc}^{n}\frac{1-p_i}{\mu_i}\prod_{\substack{j=\dacc\\j\neq i}}^{n}\frac{\mu_j}{(\mu_j-\mu_i)}\Bigg),
\label{eq:them-Cd}
\end{align}}
where $\PrBSd + \PrDd =1$.
\end{theorem}
\begin{IEEEproof}
See appendix.
\end{IEEEproof}

\subsection{Average Total Cost}
\label{sec:total-computation}

Combining Theorems \ref{thm:Cr} and \ref{thm:Cd}, one obtains the expression for $\EC=\ECr+\ECd$. Note that in general $\EC$ is not monotone with $\dr$. We can derive the following result for $\dr\rightarrow 0$ and $\dr\rightarrow \infty$.
\begin{corollary}\label{cor:Cinst}
$\lim_{\dr\to0}\EC=\frac{\rhoD}{M}(n\mu\gD+N\omega \dacc\alpha)$. Moreover, for $\mu>0$, $\lim_{\dr\to\infty}\EC=N\omega \rhoBS$.
\end{corollary}
For instantaneous repair ($\dr =0$), both repair and download are always performed from the storage nodes. Thus, the two terms in $\EC$ for $\dr\rightarrow 0$ in Corollary~\ref{cor:Cinst} correspond to the repair and download costs in the D2D regime. For $\dr \rightarrow \infty$, data is never repaired (hence, $\ECr=0$). For $\mu>0$, the number of storage nodes in the cell will become smaller than $\dacc$ at some point, and D2D download is not possible. Therefore, the average download cost is the average BS download cost.

\section{MDS and Regenerating Codes}
\label{sec:codes}

From Section \ref{sec:analysis} it can be seen that the total cost, $\EC$, depends on the DS system parameters $n$, $\dacc$, $r$, $\gD=r\beta$,  and $\gBS=\alpha$ (among others). This section describes how, in turn, these parameters depend on the $(\nc,\kc)$ erasure correcting codes used for storage. We consider as examples MDS codes \cite{Ryan2009} and regenerating codes \cite{5550492}.

\subsection{Maximum Distance Separable Codes}\label{sec:mds}

Assume the use of an $(\nc,\kc)$ MDS code for DS. Then, due to the MDS property, D2D repair and D2D download require to contact $r=\dacc=\kc$ storage nodes. Moreover, $\bMDS=\aMDS=\frac{M}{\kc}$, which means that $\gD=M$. The fact that an amount of information equal to the size of the entire file has to be retrieved to repair a single storage node is a known drawback of MDS codes \cite{5550492}.

The simplest MDS code is the $\nc$-replication scheme. In this case, each storage node stores the entire file, i.e., $\arep=M$. For the replication scheme, $r=\dacc=1$ and $\brep=M$. 

\subsection{Regenerating Codes}

A lower repair bandwidth $\gD$ (as compared to MDS codes) can be obtained by using regenerating codes \cite{5550492}, but at the expense of increasing $r$ \cite{5550492}. Two main classes of regenerating codes are covered here, minimum storage regenerating (MSR) codes and minimum bandwidth regenerating (MBR) codes. For given $n$ and $k$, MSR codes yield the best storage efficiency, i.e., $\aMSR$ is minimum, while MBR codes achieve minimum D2D repair bandwidth, i.e., $\gD$ is minimum.


For an $(\nc,\kc)$ MSR code in a DS system, $\dacc=\kc$. Moreover, $r\in\{\kc,\dots,n-1\}$ storage nodes are contacted during the D2D repair process. Hence, the download cost $\ECd$ for an $(\nc,\kc)$ MSR code is equal to the one of an $(\nc,\kc)$ MDS code. However, $\bMSR = \frac{M}{\dacc}\frac{1}{r-\dacc+1}\le \bMDS$ \cite{5550492}. $\gD=r\bMSR$ is minimized for $r=n-1$. For $r=\kc$, the total cost $\EC$ of the MSR code is equal to that of the MDS code.

As described in \cite{5550492}, an $(\nc,\kc)$ MBR code in a DS system has $r\in\{\dacc,\ldots,n-1\}$ and $\gD=r\bMBR=\frac{M}{\dacc}\frac{2r}{2r-\dacc+1}$. Furthermore, $\gD=\aMBR=\frac{M}{\kc}$\cite{5550492}, where the last equality comes from \eqref{eq:alpha}. The relationship between $\kc$, $\dacc$ and $r$ is therefore $\kc=\dacc\frac{2r-h+1}{2r}$.

\section{Numerical results}
\label{sec:results}

\begin{figure}[!t]
	\includegraphics{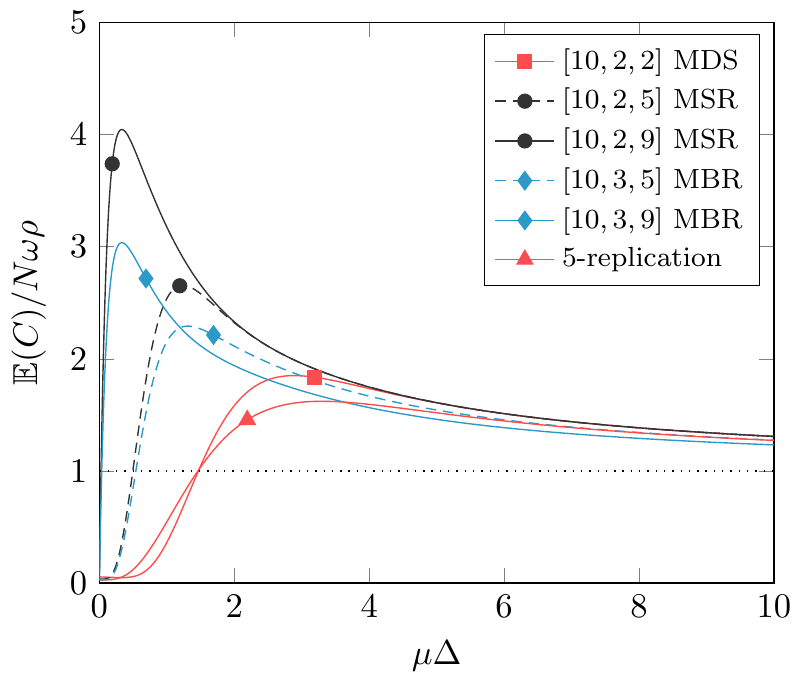}
	\vspace{-2ex}
	\caption{Normalized total cost $\EC/N\omega\rho$ versus the normalized repair interval $\mu\dr$ for MDS, MSR, and MBR codes.} 
	\label{fig:C}
	\vspace{-2ex}
\end{figure}

\begin{figure}[!t]
	\includegraphics{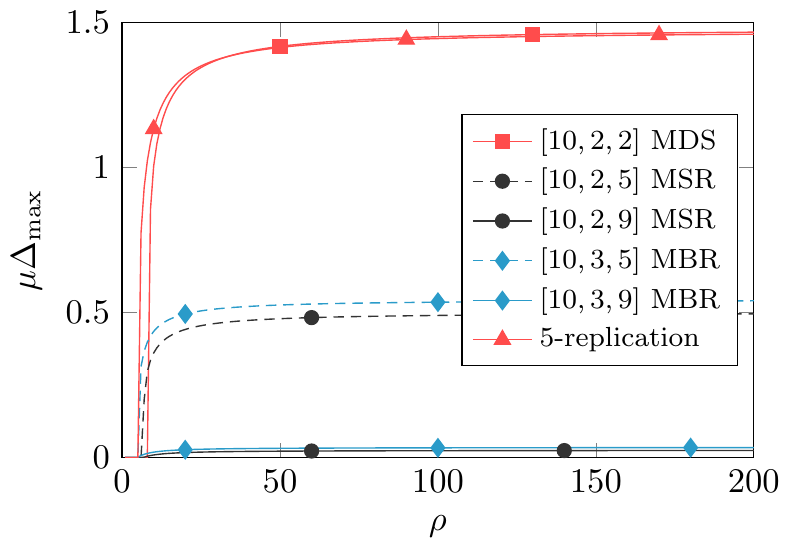}
	\vspace{-2ex}
	\caption{$\mu\drmax$ as a function of the cost ratio $\rho$.}
	\label{fig:norepair}
	\vspace{-2ex}
\end{figure}

\begin{figure}[!t]
	\centering{\subfloat[Small $\mu\dr$.]{
		\includegraphics{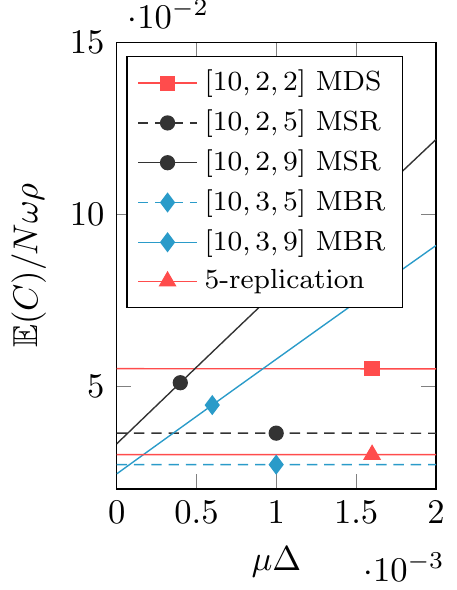}\label{fig:zoom}}
	\subfloat[Moderate $\mu\dr$.]{
		\includegraphics{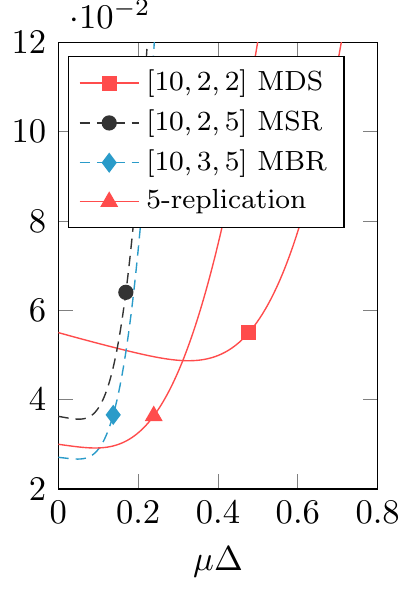}\label{fig:opt}}}
		\vspace{-2ex}
	\caption{Normalized total cost $\EC/N\omega\rho$ for the same codes as in Fig.~\ref{fig:C}, but for much smaller normalized repair intervals $\mu\dr$.}
	\label{fig:detail}
	\vspace{-2ex}
\end{figure}

In this section, we evaluate the total cost $\EC$ for MDS and regenerating codes. For the results, we consider a network with $N=100$ average nodes, request rate $\omega = 0.5$, and a cost ratio $\rho=200$. Also, the storage budget is set to $\Gamma=5$. Without loss of generality we set $\rhoD=1$ c.u./bit, i.e., $\rho=\rhoBS$. To specify a code, we use the alternative notation $[n,\dacc,r]$.

Fig.~\ref{fig:C} shows the value of the normalized cost $\EC/N\omega\rho$ versus the normalized repair interval $\mu \dr$ for $\mu = 50$, for the $[10,2,2]$ MDS code, the $[10,2,r]$ MSR code with $r\in\{5,9\}$, i.e., moderate and high repair access respectively, the $[10,3,r]$ MBR code with $r\in\{5,9\}$ and the 5-replication scheme. The code rate for all codes is $\Rc=1/5$, except for the $[10,3,5]$ MBR code that has $\Rc=6/25=0.24$ and the $[10,3,9]$ MBR code that has $\Rc=4/15\approx 0.27$. In the figure, $\mu \dr=1$ means that the repair interval is equal to one average node lifetime.

The code parameters are chosen to highlight particularly interesting behaviors of the different codes. Note that since $\alpha$, $\beta$ (and hence $\gD$) and $\gBS$ are proportional to the file size $M$, as specified in Section \ref{sec:codes}, the repair and download cost in \eqref{eq:Cr} and \eqref{eq:them-Cd}, respectively, are independent of the file size $M$. From Corollary~\ref{cor:Cinst}, $\EC/N\omega\rho\to1$ (the cost of always downloading content from the BS) when $\dr\to\infty$. We observe from Fig.~\ref{fig:C}  that this is indeed the case. It is interesting to point out that the normalized total cost exceeds $1$ for values of the repair interval larger than a threshold $\drmax$. We define the maximum repair interval as 
\begin{equation}
	\drmax\triangleq\sup\left\{\dr:\EC<\lim_{\dr\to\infty}\EC\right\}.
\end{equation}

For $\dr>\drmax$, retrieving the file from the BS is always less costly, therefore storing data in the nodes is useless. Clearly, $\dr_\text{max}$ is a function of the cost ratio $\rho$.
Fig.~\ref{fig:norepair} shows $\mu \dr_\text{max}$ as a function of $\rho\in[1,200]$, for all codes in Fig.~\ref{fig:C}. 
We observe that if $\rho<5$, approximately, it is never beneficial to use the devices for storage, i.e., the file should always be downloaded from the BS. As $\rho$ increases, storing data in the mobile devices is beneficial, if repair is performed with $\dr\le \drmax$. The regenerating codes with high repair access require very frequent repairs. Although not included here due to space constraints, the same is true for other MSR and MBR codes with high repair access. The MDS codes and the regenerating codes with moderate repair access require less frequent repairs; for large $\rho$, the repair interval must be at most around 1.5 and 0.5 average node lifetimes respectively.

For the same parameters and codes used in Fig.~\ref{fig:C}, Fig.~\ref{fig:detail} shows the normalized total cost for shorter repair intervals. We observe that instantaneous repair is optimal for the MBR and MSR codes with $r=9$ (Fig.~\ref{fig:detail}(a)). 
On the other hand, $\EC$ for the MDS codes and the regenerating codes with moderate repair access is minimized for $\dr>0$ (Fig.~\ref{fig:detail}(b)). 

\section{Conclusions}

We considered distributed storage for a wireless network where data is stored in a distributed manner across mobile devices. We introduced a repair scheduling where the repair of the data lost due to device departures is performed periodically. We derived analytical expressions for the total communication cost, due to repair and download, as a function of the repair interval. For a particular network, we showed that there exists a maximum value of the repair interval after which retrieving the file from the BS is always less costly. Therefore, DS is useful if the repair can be performed frequently enough.
Instantaneous repair is not always the best solution. The optimal repair interval that minimizes the total communication cost depends on the code used for storage. For a given repair interval, one should find the code that minimizes the total communication cost. A more thorough investigation is left for future research.

\appendices
\section{Outline of the Proof of Theorem \ref{thm:Cd}}

A file request entails a cost $\rhoD\dacc\alpha$ with probability $\PrDd$, and a cost $\rhoBS M$ with probability $\PrBSd$. The overall request rate per t.u. is $N \omega$.  Normalizing by the file size $M$ gives the first equality in \eqref{eq:them-Cd}. In the following, we prove the last equality of the theorem.

\begin{figure}[t!]
\centering
\includegraphics{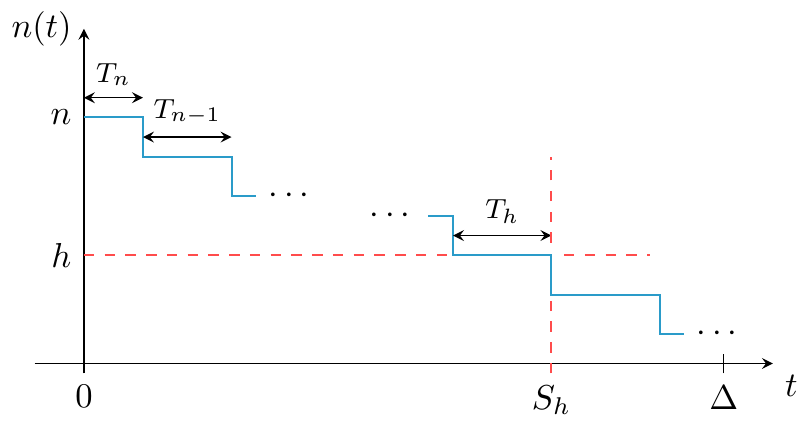}
\vspace{-2ex}
\caption{The number of available storage nodes vs. time $t$, within the repair interval $\dr$. 
At $t=0$, there are $n$ nodes available. During the intervals $T_i$, there are $i$ nodes. 
Hence, during the time interval $t \in [0, S_\dacc)$ there are at least $\dacc$ nodes available for D2D download.
}\label{fig:Ss}
\vspace{-2ex}
\end{figure}

Within a repair interval, the number of storage nodes $n(t)$ in the cell is described by a Poisson death process \cite{Miller2004}. 
Denote by $T_i$ the time interval for which $n(t)=i$, $i \in \{\dacc, \ldots, n\}$ (see Fig. \ref{fig:Ss} for illustration).
$T_i$ is exponentially distributed with rate $\mu_i=i\mu$. 
Denote by $S_\dacc$ the time instant within the repair interval at which $n(t)$ changes from $\dacc$ to $\dacc-1$. 
Then,
\begin{equation} \label{eq:Sk}
	S_\dacc = \sum_{i=\dacc}^{n}T_i.
\end{equation}

The pdf of $S_\dacc$ is given by \cite{Bolch2006}
\begin{equation}\label{eq:Sspdf}
	f_{S_\dacc}(t)=\sum_{i=\dacc}^n\frac{\mu_n\mu_{n-1}\ldots\mu_{\dacc}}{\prod_{\substack{j=\dacc\\j\neq i}}^{n}(\mu_j-\mu_i)}e^{-\mu_it},\quad t\geq0.
\end{equation}

We are interested in the distribution of file requests within a repair interval $\dr$.
Let $W_{l}$ be the time instant of the $l$th request. 
$W_l$ is computed as the sum of $l$ inter-request times with pdf given by \eqref{eq:Trpdf}. Thus,
$W_l$ is an Erlang distributed random variable with pdf \cite{Miller2004}\begin{equation}
	f_{W_l}(t)=\frac{\omega^lt^{l-1}e^{-\omega t}}{(l-1)!},\quad t\geq0.
\end{equation}
Define $\tilde W_l\triangleq W_l\mod\dr$. The following result holds.
\begin{lemma}
\label{lem:Sipdf}
The distribution of $\tilde W_l$ for $t\in[0,\dr)$ is 
$$ f_{\tilde W_l}(t) =\sum_{i=0}^{\infty}\frac{\omega^l(t+i\dr)^{l-1}e^{-\omega(t+i\dr)}}{(l-1)!}.$$
\end{lemma}
\begin{lemma}\label{lem:Sipdfb} 
$\lim_{l\rightarrow\infty}f_{\tilde W_l}(t) = \frac{1}{\dr}.$
\end{lemma} 
The proofs are omitted due to lack of space. It can be verified numerically that $f_{\tilde W_l}(t)$ converges to the uniform distribution already for small values of $l$.


D2D download is possible if at least $\dacc$ storage nodes are available in the network. 
Thus, given the sequence of random variables $\{\tilde W_1, \tilde W_2, \ldots\}$, 
\begin{align*}
\PrDd & = \lim_{L \rightarrow \infty}\frac{1}{L}\sum_{l=1}^{L}\Pr(\tilde W_l <S_\dacc)\\
&\approx \Pr(\tilde W_{\infty} <S_\dacc),
\end{align*}
where the approximation follows because for large enough $l$, $f_{\tilde W_l}(t) \approx \frac{1}{\dr}$.

Now, using \eqref{eq:Sspdf}, after some calculations we obtain
\begin{align}
\PrDd	=\frac{1}{\dr}\sum_{i=\dacc}^{n}\frac{1-p_i}{\mu_i}\prod_{\substack{j=\dacc\\j\neq i}}^{n}\frac{\mu_j}{(\mu_j-\mu_i)}.
\label{eq:prdd}
	\end{align}

Finally, using \eqref{eq:prdd} and $\PrBSd = 1- \PrDd$ we obtain \eqref{eq:them-Cd}. This completes the proof.

\bibliographystyle{IEEEtran}
\bibliography{library}



\end{document}